\documentclass[preprint,aps,prd,showpacs,superscriptaddress,nofootinbib,tightenlines]{revtex4}

\usepackage{amsfonts}
\usepackage{mathrsfs}
\usepackage{graphicx}
\usepackage{amsmath}
\usepackage{amssymb}
\usepackage{bm}
\usepackage{bbm}
\usepackage{epsfig}
\usepackage{multirow}
\usepackage{float}
\usepackage{array}
\usepackage{tabularx}
\usepackage{xcolor}

\makeatletter
\def\hlinew#1{%
  \noalign{\ifnum0=`}\fi\hrule \@height #1 \futurelet
   \reserved@a\@xhline}
\usepackage{array}
\newcommand{\PreserveBackslash}[1]{\let\temp=\\#1\let\\=\temp}
\newcolumntype{C}[1]{>{\PreserveBackslash\centering}p{#1}}
\newcolumntype{R}[1]{>{\PreserveBackslash\raggedleft}p{#1}}
\newcolumntype{L}[1]{>{\PreserveBackslash\raggedright}p{#1}}
\usepackage{float}

\def\OMIT#1{}

\newcommand{\nn}{\nonumber}

\newcommand{\bqa}{\begin{eqnarray}}
\newcommand{\eqa}{\end{eqnarray}}

\begin{document}

\title{
$\bm{\mathcal O}\bm{(}\bm{\alpha}_{\bm s}\bm{)}$ corrections to
$\bm{e}^{\bm{+}}
\bm{e}^{\bm{-}}\bm{\rightarrow}\bm{J}\bm{/}\bm{\psi}\bm{+}\bm{\eta}_{\bm{c2}}
\bm{(}\bm{\chi}^{\bm \prime}_{\bm{c}\bm{1}}\bm{)}$ at $\bm{B}$
factories}


\author{Hai-Rong Dong\footnote{E-mail: donghr@ihep.ac.cn}}
\affiliation{Institute of High Energy Physics, Chinese Academy of
Sciences, Beijing 100049, China\vspace{0.2cm}}

\author{Feng Feng\footnote{E-mail: fengf@ihep.ac.cn}}
\affiliation{Center for High Energy Physics, Peking University,
Beijing 100871, China\vspace{0.2cm}}

\author{Yu Jia\footnote{E-mail: jiay@ihep.ac.cn}}
\affiliation{Institute of High Energy Physics, Chinese Academy of
Sciences, Beijing 100049, China\vspace{0.2cm}}
\affiliation{Theoretical Physics Center for Science Facilities,
Chinese Academy of Sciences, Beijing 100049, China\vspace{0.2cm}}

\date{\today}
\begin{abstract}

We investigate the ${\cal O}(\alpha_s)$ correction to $e^+e^-\to
J/\psi+\eta_{c2}$  in the NRQCD factorization approach. A detailed
comparative study between $e^+e^-\to J/\psi+\eta_{c2}$ and
$e^+e^-\to J/\psi+\chi^\prime_{c1}$ at $B$ factory energy is also
carried out. After incorporating the ${\cal O}(\alpha_s)$
correction, we predict the cross section for the former process to
be around 0.3 fb, while that of the latter about 6 times greater.
The outgoing $J/\psi$ is found to be dominantly
transversely-polarized in the former process, while
longitudinally-polarized in the latter. These features may provide
valuable guidance for the future experiment to examine the ${}^3P_1$
or ${}^1D_2$ charmonium option of the $X(3872)$ meson through the
exclusive double-charmonium production processes. The observation
potential of $e^+e^-\to J/\psi+\chi^\prime_{c1}$ looks bright for
the current data sample of the \textsc{Belle} experiment, provided
that the $\chi^\prime_{c1}$ is indeed the narrow $X(3872)$ state. In
the appendix, we also identify the coefficients of the double
logarithms of form $\ln^2(s/m_c^2)$ associated with all the relevant
next-to-leading order Feynman diagrams, for the helicity-suppressed
double-charmonium production channels $e^+ e^- \to J/\psi+
\eta_{c2}(\eta_c,\chi_{c0,1,2})$.
\\


\end{abstract}

\maketitle

\newpage

\section{Introduction}
\label{Introduction}

Since a number of double charmonium production processes were first
discovered at $B$ factory a decade
ago~\cite{Abe:2002rb,Abe:2004ww,Aubert:2005tj}, this topic has
spurred a widespread interest~\cite{Brambilla:2010cs}. Firstly, this
production environment provides a unique and powerful means to
search for the new $C$-even charmonium states, especially those $X$,
$Y$, $Z$ states, by fitting the recoil mass spectrum against the
$J/\psi$ ($\psi'$). The most famous examples are the discovery of
the $X(3940)$~\cite{Abe:2007jn} and the $X(4160)$~\cite{Abe:2007sya}
by this way.

Another appealing reason to study double charmonium production is
that it provides a new stage to sharpen our understanding toward
perturbative QCD, especially toward the application of the
light-cone approach~\cite{Lepage:1980fj,Chernyak:1983ej} and the
nonrelativistic QCD (NRQCD) factorization
approach~\cite{Bodwin:1994jh} to hard exclusive reactions involving
heavy quarkonium.

The most famous double-charmonium production process is perhaps
$e^+e^-\to J/\psi+\eta_c$. The original lowest order (LO) NRQCD
predictions to this process~\cite{Braaten:2002fi,Liu:2002wq} is
about one order of magnitude smaller than the
measurement~\cite{Abe:2002rb}. This acute discrepancy has triggered
a great amount of theoretical investigations in both NRQCD and
light-cone approaches~\cite{Hagiwara:2003cw,Ma:2004qf,Bondar:2004sv,
Bodwin:2006dm,Zhang:2005cha,Gong:2007db, He:2007te, Bodwin:2007ga,
Braguta:2008tg,Dong:2012xx,Li:2013qp}. One crucial element in
alleviating the discrepancy between the NRQCD prediction and the
data is the significant and positive next-to-leading order (NLO)
perturbative corrections~\cite{Zhang:2005cha,Gong:2007db}. By
contrast, owing to some long-standing theoretical obstacles, the NLO
correction to this helicity-suppressed process in the light-cone
approach has never been successfully worked out. As a consequence,
despite some shortcomings, the NRQCD approach seems to be the only
viable method which is based on the first principle and also
systematically improvable.

Recently, the ${\cal O}(\alpha_s)$ corrections to the
double-charmonium production processes $e^+e^-\to
J/\psi+\chi_{c0,1,2}$ have also been investigated in NRQCD
approach~\cite{Wang:2011qg,Dong:2011fb}. For some production
channels, the effect of the NLO perturbative corrections can be
important.

In this work, we plan to carry out a comprehensive study of the NLO
perturbative corrections to the processes $e^+e^-\to
J/\psi+\eta_{c2}$ and $e^+e^-\to J/\psi+\chi^\prime_{c1}$, again in
the NRQCD factorization framework. The calculation for the former
process is new, while that for the latter can be readily adapted
from Ref.~\cite{Dong:2011fb}. This work should be considered as a
sequel of Refs.~\cite{Dong:2011fb,Dong:2012xx}.

Our interest in conducting such a comparative study is strongly
motivated by the controversy recently raised at the quantum number
of the $X(3872)$ meson, whether it being $1^{++}$ or
$2^{-+}$~\cite{delAmoSanchez:2010jr}. Consequently, the canonical
charmonium options for the $X(3872)$ would be the $\chi'_{c1}$ and
$\eta_{c2}$, respectively. In literature, there have already existed
a number of studies about critically distinguishing the properties
of $\chi'_{c1}$ and $\eta_{c2}$, such as the radiative transitions
$\eta_{c2}(\chi'_{c1})\to
J/\psi(\psi^\prime)$~\cite{Jia:2010jn,Kalashnikova:2010hv,Yang:2012mya},
the $\eta_{c2}(\chi'_{c1})$ hadroproduction
rates~\cite{Burns:2010qq}, or $\eta_{c2}(\chi'_{c1})$ production
rates in $B$ decay~\cite{Fan:2011aa}. By confronting the established
properties of $X(3872)$, all these studies tend to disfavor the
${}^1D_2$ assignment of the $X(3872)$ meson. We hope that
double-charmonium production can also be added to the above list as
a valuable means to help clarify the situation.

It turns out that the production rate of $e^+e^-\to
J/\psi+\chi'_{c1}$ is about 6-7 times greater than that of
$e^+e^-\to J/\psi+\eta_{c2}$. Based on the 1 ${\rm ab}^{-1}$ data
currently accumulated at the \textsc{Belle} experiment, it looks
promising to observe the former process, if the $\chi^\prime_{c1}$
can indeed be regarded as the $X(3872)$ particle with very narrow
width. This seems to constitute the strong enough incentive for
experimentalists to make an updated analysis of the double
charmonium production at $B$ factory.

The rest of the paper is organized as follows.
In Section~\ref{hsr:red:hel:ampl}, we specify the helicity selection
rule suited for the hard exclusive reaction $e^+e^-\rightarrow
J/\psi+\eta_{c2}$, and give the definition for the dimensionless,
reduced helicity amplitudes.
In Section~\ref{LO:result}, we present the leading order expressions
for all the independent helicity amplitudes in the NRQCD
factorization framework.
In Section~\ref{NLO:result}, we first review some key technical
issues about the ${\cal O}(\alpha_s)$ calculation, then present the
asymptotic expressions for the NLO perturbative corrections to all
the encountered helicity amplitudes. The pattern of the
double-logarithmic scaling violation is confirmed once again.
In Section~\ref{phenomenology}, a comparative study is performed for
both unpolarized and polarized cross sections between the processes
$e^- e^+ \to J/\psi+\eta_{c2}$ and $e^- e^+ \to
J/\psi+\chi^\prime_{c1}$. This study may shed some light on
unveiling the quantum number of the $X(3872)$ meson in the future
double-charmonium production experiments.
Finally we summarize in Section~\ref{summary}.
In Appendix~\ref{double:log:anatomy}, we tabulate the coefficients
of the double logarithm $\ln^2(s/m_c^2)$ associated with all the
relevant NLO Feynman diagrams, for all the double-charmonium
production channels we have studied so far, {\it i.e.}, $e^+ e^- \to
J/\psi+ \eta_{c2}(\eta_c,\chi_{c0,1,2})$.

\section{Helicity selection rule and reduced helicity amplitudes}
\label{hsr:red:hel:ampl}

It is often desirable to glean more information than simply present
the unpolarized cross section for a hard exclusive reaction,
especially for the double-charmonium production process considered
in this work. It is of particular advantage to making explicit
predictions to various $J/\psi+\eta_{c2}$ production rates for
different helicity configurations. The underlying reasons of
carrying out such detailed studies are two-fold. On the experimental
ground, once the sufficient statistics is achieved, the helicity
amplitudes themselves in principle can be measured by studying the
angular distributions of these charmonia and their decay products;
from the theoretical perspective, it is also instructive to stay
with the helicity amplitudes. The reason is that, for a hard
exclusive reaction, the relative importance of the polarized cross
section in a given helicity channel is dictated by the celebrated
{\it helicity selection rule} (HSR)~\cite{Brodsky:1981kj}.

We are interested in the hard-scattering limit $\sqrt{s}\gg m_c\gg
\Lambda_{\rm QCD}$, where $\sqrt{s}$ stands for the center-of-mass
energy of the $e^+e^-$ collider, $m_c$ for the charm quark mass, and
$\Lambda_{\rm QCD}$ for the intrinsic QCD scale. In this limit, the
asymptotic behavior of the production rate for $J/\psi+\eta_{c2}$ in
a definite helicity configuration follows from the
HSR~\cite{Braaten:2002fi}:
\bqa
{\sigma[e^+e^-\rightarrow J/\psi(\lambda_1) +
\eta_{c2}(\lambda_2)]\over \sigma[e^+e^-\to \mu^+\mu^-]} & \sim &
v^{10} \left({m_c^2\over s}\right)^{2+|\lambda_1+\lambda_2|},
\label{helicity:selection:rule}
\eqa
where $\lambda_1$, $\lambda_2$ represent the helicities carried by
the $J/\psi$, $\eta_{c2}$, respectively. $v$ denotes the
characteristic velocity of charm quark inside a charmonium. Equation
(\ref{helicity:selection:rule}) implies that the helicity state
which exhibits the slowest asymptotic decrease, thus constitutes the
``leading-twist" contribution, {\it i.e.}, $\sigma\sim 1/s^3$, is
$(\lambda_1,\lambda_2)=(0,0)$.

We have chosen by default to work in the $e^+e^-$ center-of-mass
frame. Let $|{\bf P}|$ signify the magnitude of the momentum carried
by the $J/\psi$ ($\eta_{c2}$), and $\theta$ denote the angle between
the moving directions of the $J/\psi$ and the $e^-$ beam. Following
the steps elaborated in \cite{Dong:2011fb}, the differential rate
for polarized $J/\psi+\eta_{c2}$ production in $e^+e^-$ annihilation
can be expressed as
\bqa
& & {d\sigma[e^+e^-\to J/\psi(\lambda_1)+\eta_{c2}(\lambda_2)]\over
d\cos \theta}
\nn\\
&= & {\alpha \over 8 s^2} \left({|{\bf P}|\over \sqrt{s}}\right)
|{\mathcal A}_{\lambda_1,\lambda_2}|^2 \times \Big\{
\begin{array}{c}
{1+\cos^2\theta\over 2}\qquad\qquad(\lambda_1-\lambda_2={\pm 1})
\\
\sin^2\theta\qquad\qquad (\lambda_1-\lambda_2=0).
\end{array}
\label{polar:diff:cross:section}
\eqa
where ${\mathcal A}_{\lambda_1,\lambda_2}$ is the helicity amplitude
associated with the virtual photon decay into $J/\psi+\eta_{c2}$
carrying the helicity component $(\lambda_1,\lambda_2)$.

Parity invariance can be invoked to reduce the number of independent
helicity amplitudes:
\bqa
{\mathcal A}_{\lambda_1,\lambda_2}=- {\mathcal
A}_{-\lambda_1,-\lambda_2},
\label{parity:trans:hel:ampl}
\eqa
hence the two helicity amplitudes related by flipping the helicities
of two chamonia bear the equal magnitude. An immediate consequence
of (\ref{parity:trans:hel:ampl}) is that, the virtual photon decay
into the longitudinally-polarized $J/\psi$ and $\eta_{c2}$ is
strictly forbidden by parity invariance.

Integrating (\ref{polar:diff:cross:section}) over the polar angle
$\theta$ and including all the allowed helicity states, it is then
straightforward to obtain the unpolarized cross section:
\bqa
&& \sigma [e^+e^-\rightarrow J/\psi + \eta_{c2}] = {\alpha
\sqrt{1-4r} \over 12 s^2} \left( 2\left|{\mathcal A}_{0,1}\right|^2
+ 2\left|{\mathcal A}_{1,0}\right|^2 + 2\left|{\mathcal
A}_{1,1}\right|^2 + 2\left | {\mathcal A}_{1,2}\right|^2
\right),
\label{unpol:cross:section:Jpsi:etac2}
\eqa
In conformity with the constraint $|\lambda_1-\lambda_2|\le 1$, as
demanded by angular momentum conservation, there are totally 4
independent helicity amplitudes for $\gamma^* \to J/\psi+\eta_{c2}$
(Recall that ${\mathcal A}_{0,0}=0$ owing to parity invariance). In
equation (\ref{unpol:cross:section:Jpsi:etac2}), we have also
retained a factor of 2 explicitly  to account for the contributions
from those helicity-flipped states. We have also adopted the
approximation ${2 |{\bf P}|\over \sqrt{s}}\approx \sqrt{1-4r}$ by
assuming $M_{J/\psi}\approx M_{\eta_{c2}}\approx 2m_c$.

In the NRQCD factorization framework, the product of two
nonperturbative factors, {\it i.e.}, the (second derivative of) wave
functions at the origin for the charmonia $J/\psi$, and $\eta_{c2}$:
$R_{J/\psi}(0)$, and $R_{\eta_{c2}}''(0)$, ubiquitously enters every
helicity amplitude, thereby it appears convenient to define a {\it
reduced} dimensionless helicity amplitude, of which these
nonperturbative factors are pulled out. We introduce the reduced
helicity amplitude, $a_{\lambda_1,\lambda_2}$, which is related to
the standard helicity amplitude ${\mathcal A}_{\lambda_1,\lambda_2}$
as follows:
\bqa
{\mathcal A}_{\lambda_1,\lambda_2}  &=&  { 2^8 \sqrt{5}\, e_c e \,
\alpha_s  R_{J/\psi}(0) R^{''}_{\eta_{c2}}(0)\over 3 s
m_c^2}\,r^{{1\over 2}(|\lambda_1+\lambda_2|-1)}
\,a_{\lambda_1,\lambda_2}(r),
\label{reduced:helicity:ampl:definition}
\eqa
where $e_c e={2\over 3}e$ is the electric charge of the charm quark,
$r\equiv 4m_c^2/s$ signifies a dimensionless mass ratio. To make the
HSR manifest, we have explicitly stripped off a factor $r^{{1\over
2}(|\lambda_1+\lambda_2|-1)}$ in
(\ref{reduced:helicity:ampl:definition}), so that the dimensionless
helicity amplitude $a_{\lambda_1,\lambda_2}$ is expected to scale as
${\cal O}(r^0)$.

Plugging (\ref{reduced:helicity:ampl:definition}) back into
(\ref{unpol:cross:section:Jpsi:etac2}), we obtain the NRQCD
predictions to the polarized production rate for the
$J/\psi(\lambda_1)+\eta_{c2}(\lambda_2)$ state:
\bqa
& & \sigma[e^+e^-\to J/\psi(\lambda_1)+\eta_{c2}(\lambda_2)]
\nn\\
& = & { 5\cdot 2^{16} \pi e_c^2 \alpha^2 \alpha_s^2 \over 3^3 s^4
m_c^4} \,R^2_{J/\psi}(0)
R''^2_{\eta_{c2}}(0)\,r^{|\lambda_1+\lambda_2|-1} \,(1-4r)^{1\over
2} \left| a_{\lambda_1,\lambda_2}\right|^2.
\label{pol:cross:section:from:red:hel:ampl}
\eqa

\section{LO predictions for the helicity amplitudes}
\label{LO:result}

\begin{figure}[t]
\begin{center}
\includegraphics[scale=0.7]{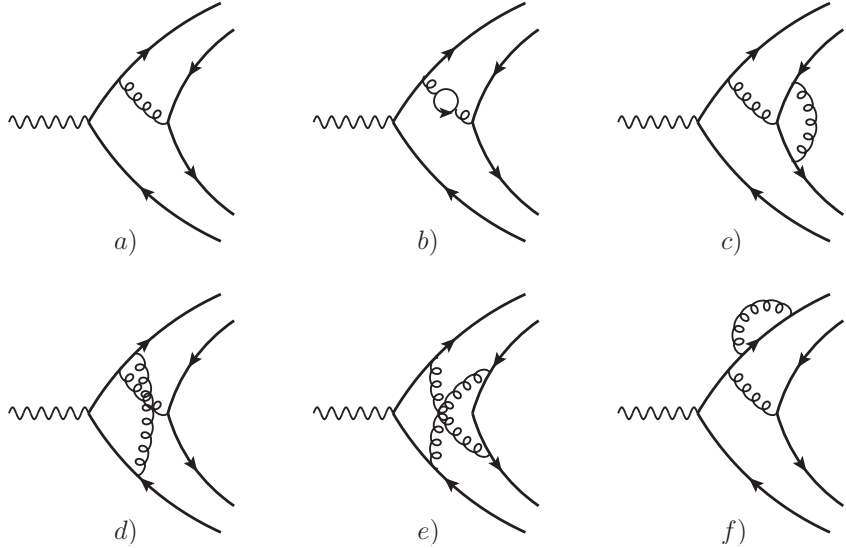}
\caption{One sample LO diagram and five sample NLO diagrams that
contribute to $\gamma^*\to J/\psi+\eta_{c2}$.
\label{feynman:diagrams}}
\end{center}
\end{figure}

The reduced helicity amplitude $a_{\lambda_1,\lambda_2}$ can be
viewed as the NRQCD short-distance coefficient, which encodes the
contribution solely stemming from the momentum region between $m_c$
and $\sqrt{s}$, so can be computed reliably in perturbation theory.
It is convenient to parameterize it as
\bqa
& & a_{\lambda_1,\lambda_2} =
a^{(0)}_{\lambda_1,\lambda_2}+{\alpha_s\over\pi}
a^{(1)}_{\lambda_1,\lambda_2}.
\eqa
\label{red:hel:ampl:organization:in:alphas}
Our central task in this work is to decipher the ${\cal
O}(\alpha_s)$ correction to the reduced amplitude,
$a^{(1)}_{\lambda_1,\lambda_2}$.

First we recapitulate the LO calculation. To proceed, it it most
convenient to consider the quark amplitude $\gamma^*\to c\bar{c}
(^3S_1^{(1)})+c\bar{c} (^1D_2^{(1)})$ using the covariant projection
technique~\cite{Bodwin:2002hg,Braaten:2002fi}. At LO in $\alpha_s$,
there are only four Feynman diagrams that contribute, one of which
is depicted in Fig.~\ref{feynman:diagrams}$a)$ (For simplicity, we
have neglected the QED fragmentation diagrams, whose effect appears
to be modest).

It is straightforward to project out 4 corresponding helicity
amplitudes from the decay amplitude for $\gamma^*\rightarrow
J/\psi+\eta_{c2}$. One then follows equation
(\ref{reduced:helicity:ampl:definition}) to read off each of the LO
reduced helicity amplitudes:
\bqa
 a^{(0)}_{\lambda_1,\lambda_2}
&=& \Big\{
\begin{array}{c}
\pm (1-4r)^{3\over 2}\qquad\qquad(\lambda_1,\lambda_2)=({\pm 1},0)
\\
0\qquad\qquad\qquad\qquad\;(\lambda_1,\lambda_2)= {\rm others}.
\end{array}
\label{red:ampl:LO:in:alphas}
\eqa
For some accidental reason, only the $J/\psi(\pm 1)+\eta_{c2}(0)$
channel has the non-vanishing cross section at LO in $\alpha_s$.
Substituting (\ref{red:ampl:LO:in:alphas}) into
(\ref{pol:cross:section:from:red:hel:ampl}), we find agreement with
the QCD part of the LO prediction for the unpolarized cross section
first given in Ref.~\cite{Braaten:2002fi}.

\section{NLO perturbative corrections to the helicity amplitudes}
\label{NLO:result}

We start this section by first sketching some technical issues about
the NLO perturbative calculations, followed by presenting the
asymptotic expressions of the ${\cal O}(\alpha_s)$ corrections to
all the reduced helicity amplitudes.

\subsection{Outline of the calculation}

At NLO in $\alpha_s$, there are 20 two-point, 20 three-point, 18
four-point, and 6 five-point one-loop diagrams for the process
$\gamma^*\to c\bar{c} (^3S_1^{(1)})+c\bar{c} (^1D_2^{(1)})$, some of
which have been illustrated in Fig.~\ref{feynman:diagrams}. The
calculation is quite similar to our preceding works on double
charmonium exclusive production, {\it i.e.}, $e^+e^-\to
J/\psi+\chi_{c0,1,2}$~\cite{Dong:2011fb} and the $e^+e^-\to
J/\psi+\eta_c$~\cite{Dong:2012xx}, so here we will only present a
very brief description.

We adopt dimensional regularization to regularize both UV and IR
singularities. We follow the 't Hooft-Veltman prescription for
$\gamma_5$, as detailed in Ref.~\cite{Jia:2011ah}. In projecting out
the $D$-wave orbital angular momentum state, one needs expand the
relative quark momentum $q$ to the quadratic order. Our strategy is
to follow the method of region~\cite{Beneke:1997zp} via directly
deducing the NRQCD short-distance coefficients, rather than
resorting to the much more expensive matching calculation, through
making the expansion in $q$ before carrying out the loop
integration.

The only nontrivial technical problem is that one may encounter some
unusual one-loop integrals which generally contain the propagators
of cubic power, as a consequence of taking the derivative over $q$
twice. The \textsc{Mathematica} package
\textsc{FIRE}~\cite{Smirnov:2008iw} and the code
\textsc{Apart}~\cite{Feng:2012iq} are utilized to reduce these
unconventional higher-point one-loop tensor integrals into a set of
masters integrals. With the aid of the integration-by-part algorithm
and partial fractioning technique, it turns out that all the
encountered master integrals are nothing but the standard 2-point
and 3-point one-loop scalar integrals, whose analytic expressions
have already been tabulated in Ref.~\cite{Gong:2007db}.

When adding the contributions of all the diagrams, and after
renormalizing the charm quark mass and the QCD coupling constant, we
finally end up with both UV and IR finite NLO expressions for the
decay amplitude $\gamma^*\to c\bar{c} (^3S_1^{(1)})+c\bar{c}
(^1D_2^{(1)})$. Since everything becomes finite, we can safely
return to the 4 spacetime dimensions, and readily project out all
the required helicity amplitudes.

In Ref.~\cite{Bodwin:2008nf}, an all-order-in-$\alpha_s$ proof for
exclusive quarkonium production has been outlined in the NRQCD
factorization context. It argues that at lowest order in $v$ and to
all orders in $\alpha_s$, NRQCD factorization holds for the
exclusive production of a $S$-wave quarkonium plus any
higher-orbital-angular-momentum quarkonium in $e^+e^-$ annihilation.
Our explicit calculation confirms that the NRQCD short-distance
coefficients affiliated with $S$-wave plus $D$-wave charmonia
production are indeed IR finite at NLO in $\alpha_s$, which is
compatible with what is asserted in \cite{Bodwin:2008nf}.

\subsection{Analytic expressions of NLO helicity amplitudes}

\begin{figure}[tbH]
\begin{center}
\includegraphics[scale=0.55]{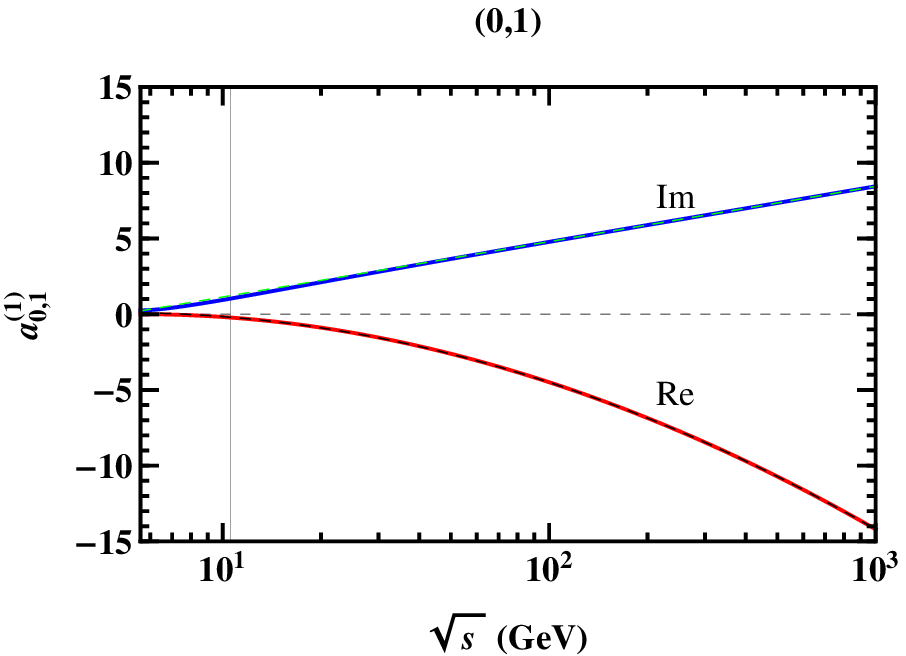}
\includegraphics[scale=0.55]{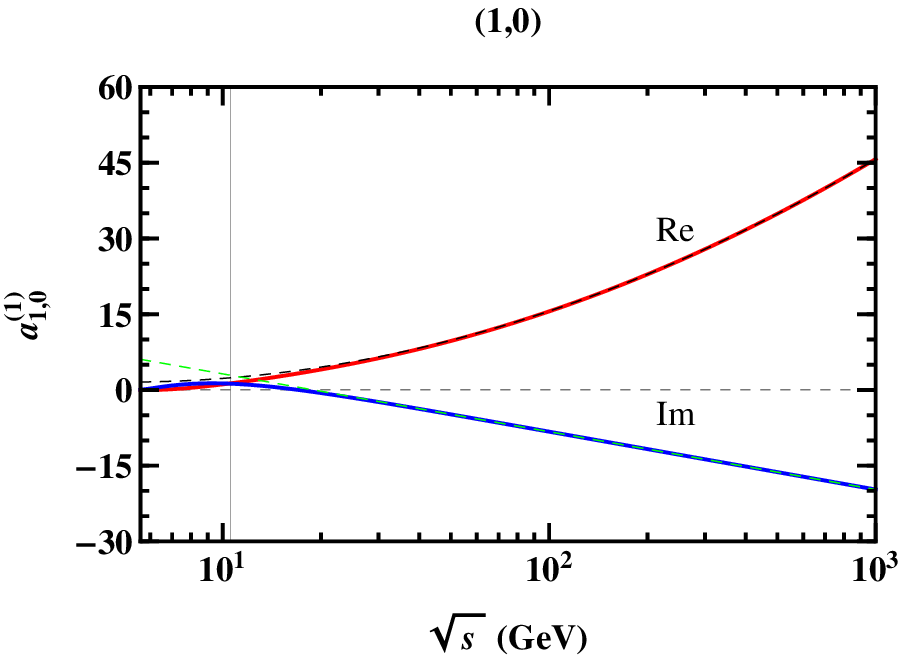}
\includegraphics[scale=0.55]{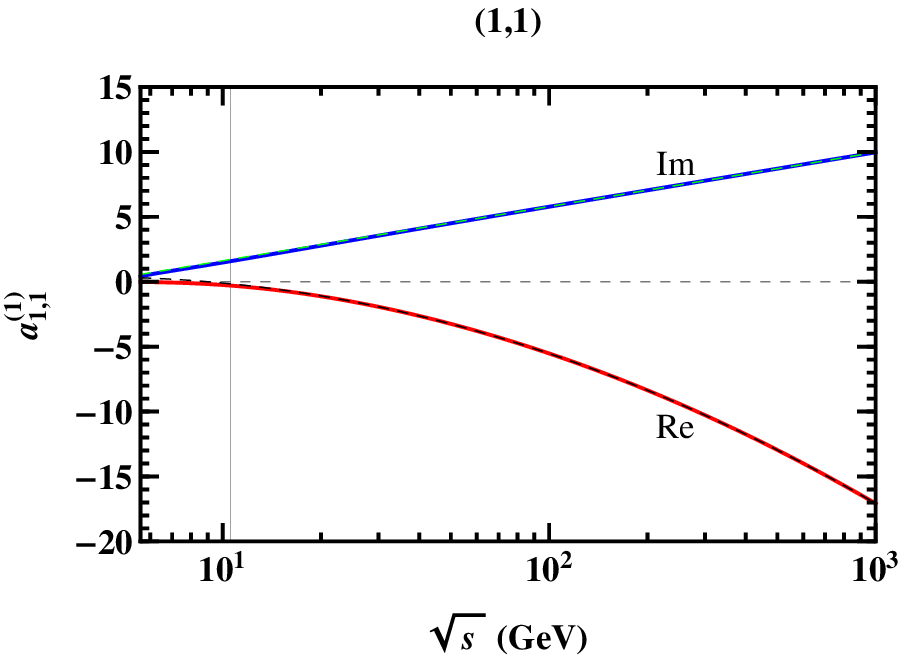}
\includegraphics[scale=0.55]{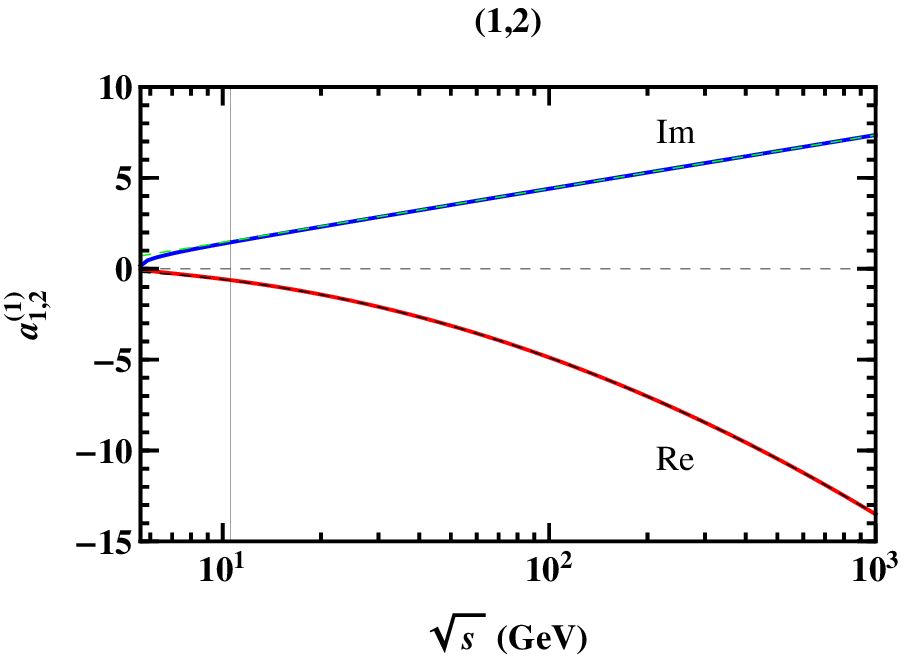}
\caption{Variation of the ${\cal O}(\alpha_s)$ reduced helicity
amplitudes $a^{(1)}_{\lambda_1,\lambda_2}$ with $\sqrt{s}$. We take
$\mu={\sqrt{s}\over 2}$ and $m_c=1.4$ GeV. The solid curves
correspond to the exact NLO results, and the dashed curves represent
the asymptotic ones as given in
(\ref{red:ampl:asymp:NLO:Jpsi:etac2}). The vertical mark is placed
at the $B$ factory energy $\sqrt{s} = 10.58$ GeV.
\label{a:functions:NLO:jpsi:etac2} }
\end{center}
\end{figure}

The $a^{(1)}_{\lambda_1,\lambda_2}$ are complex-valued, whose
analytic expressions are in general quite lengthy. Rather than
reproducing their cumbersome-looking expressions here, we are
content with presenting their numerical values over a wide range of
$r$, as shown in Fig.~\ref{a:functions:NLO:jpsi:etac2}.

As a matter of fact, it seems much more illuminating to know the
asymptotic behaviors of the helicity amplitudes in the limit
$\sqrt{s}\gg m_c$. At NLO in $\alpha_s$, one anticipates to see the
logarithmic scaling violation to the naive power-law HSR as
indicated in (\ref{helicity:selection:rule}). Furthermore,
conducting the asymptotic expansion in NRQCD short-distance
coefficients is theoretically appealing, since it is equivalent to
disentangling the contributions occurring at the ``hard" scale
(virtuality $\sim s$) from the ``lower-energy" collinear/soft
sectors (virtuality $\sim m_c^2$), by which one can intimately link
the NRQCD factorization approach and the light-cone
approach~\cite{Shifman:1980dk,Ma:2006hc,Bell:2008er,Jia:2008ep,Jia:2010fw}.
Such an asymptotic expansion on the NRQCD hard coefficients has been
carried out for a number of exclusive double-quarkonium production
processes, either in $e^+e^-$
annihilation~\cite{Jia:2010fw,Dong:2011fb,Dong:2012xx}, or from
bottomonium
decay~\cite{Jia:2007hy,Gong:2008ue,Jia:2010fw,Xu:2012uh}, and some
general pattern about logarithmic scaling violation has been
recognized~\cite{Jia:2010fw,Dong:2011fb}.

According to the definition in
(\ref{pol:cross:section:from:red:hel:ampl}), we find the asymptotic
expressions of  four reduced NLO helicity amplitudes for $\gamma^*
\rightarrow J/\psi+\eta_{c2}$ to be
\begin{subequations}
\bqa
& & a^{(1)}_{0,1}\Big|_{\rm asym} = -{\sqrt{3}\over 96} \Bigg\{ 7
\ln^2 r - 2(22-43\ln 2) \ln r - {286\over 3} - 12\pi^2 + {1024\over
3}\ln2 - 41\ln^2 2
\nn\\
&& + 2 i \pi \left(7 \ln r - 22 +43 \ln 2 \right) \Bigg\} ,
\\
& & a^{(1)}_{1,0}\Big|_{\rm asym} = {1\over 48} \Bigg\{19 \ln^2 r
+19 (1+2\ln2) \ln r + 12 \beta_0 \bigg(\ln{4\mu^2 \over s} + {8\over
3}\bigg)-{478\over 3} + {244\over 3}\ln2
\nn\\
&& -129\ln^2 2 + i\pi \bigg( 38\ln r + 12 \beta_0 + 19+38\ln2
\bigg)\Bigg\},
\\
& & a^{(1)}_{1,1}\Big|_{\rm asym} = -{\sqrt{3}\over 24} \Bigg\{  2
\ln^2 r + (19-23\ln 2) \ln r - {16\over 3} + {9\pi^2\over
2}-{143\over 3}\ln2- {41\over 2}\ln^2 2
\nn\\
&& + i \pi \left(4 \ln r+ 19 -23 \ln 2 \right) \Bigg\},
\\
& & a^{(1)}_{1,2}\Big|_{\rm asym} = -{\sqrt{6}\over 24} \Bigg\{
\ln^2 r -7 (2-3\ln 2) \ln r - {15\over 2} -{19\pi^2\over 6} + 51
\ln2 - {17 \over 2}\ln^2 2
\nn\\
&& + i \pi \left(2 \ln r-14+21\ln 2 \right) \Bigg\},
\eqa
\label{red:ampl:asymp:NLO:Jpsi:etac2}
\end{subequations}
where $\mu$ is the renormalization scale, and $\beta_0={11\over
3}C_A-{2\over 3}n_f$ is the one-loop coefficient of the QCD $\beta$
function, and $n_f=4$ denotes the number of active quark flavors. In
contrast with (\ref{red:ampl:LO:in:alphas}), all the four reduced
helicity amplitudes receive non-vanishing ${\cal O}(\alpha_s)$
corrections.

Note that the $\beta_0\ln(4\mu^2/s)$ term only resides in the $(\pm
1,0)$ channel, since this is the only channel that has a
non-vanishing tree-level amplitude. For the sake of comparison, all
these asymptotic results of the reduced helicity amplitudes are also
shown in Fig.~\ref{a:functions:NLO:jpsi:etac2}, in juxtapose with
the corresponding exact NLO results. These asymptotic results appear
to converge with the exact ones decently well even at relatively
lower $\sqrt{s}$, say, at $B$ factory energy.

From Eq.~(\ref{red:ampl:asymp:NLO:Jpsi:etac2}), one sees that the
leading scaling violation is due to the double logarithm $\ln^2 r$.
This constitutes another example to further corroborate the earlier
conjecture: the occurrence of $\ln^2 r$ in the one-loop NRQCD
short-distance coefficients is always affiliated with the
helicity-suppressed channels in exclusive double charmonium
production processes~\cite{Jia:2010fw,Dong:2011fb}.

\section{Phenomenology}
\label{phenomenology}

Aside from a comprehensive analysis on the process $e^+e^-\to J/\psi
+ \eta_{c2}$ at the $B$ factory, in this section we also target at a
detailed investigation on the process $e^+e^-\to J/\psi +
\chi'_{c1}$, where the essential elements have already been set up
in \cite{Dong:2011fb}. This study is largely motivated by the recent
concern about the nature of the $X(3872)$ meson, whether its quantum
number is $1^{++}$ or $2^{-+}$. The canonical charmonium options for
the $X(3872)$ are the $\chi'_{c1}$ and $\eta_{c2}$, respectively.
With this consideration in mind, our study may provide some useful
guidance on unveiling the quantum number of the $X(3872)$ through
the dedicated double-charmonium experiment at future $B$ factory.

Before making concrete predictions to the production rates for
$e^+e^-\to J/\psi + \eta_{c2}$, we first address a subtlety about
utilizing (\ref{pol:cross:section:from:red:hel:ampl}). For the
$J/\psi(\pm 1) + \eta_{c2}(0)$ helicity state, the canonic way of
identifying the ${\cal O}(\alpha_s)$ correction is by replacing
$\left| a_{\pm 1,0}\right|^2$ with $2{\alpha_s\over
\pi}\Re[a^{(0)}_{\pm 1,0} a^{(1)}_{\pm 1,0}]$. Because of the null
tree-level amplitudes for the remaining helicity configurations, the
${\cal O}(\alpha_s)$ correction to the total cross section is solely
from the $(\pm 1,0)$ channels. On the other hand, the polarized
cross section for each helicity state is also a physical observable
by itself. Since the non-vanishing amplitudes are first generated at
${\cal O}(\alpha_s)$ for the remaining helicity states, it is
thereby also sensible to interpret $\left|
a_{\lambda_1,\lambda_2}\right|^2=\left({\alpha_s\over \pi}\right)^2
\left| a^{(1)}_{\lambda_1,\lambda_2}\right|^2$ for
$(\lambda_1,\lambda_2)\neq (\pm 1,0)$. Though formally of order
$\alpha_s^2$, these new pieces do constitute the leading
contributions to the respective polarized cross sections, thereby
still being consistent~\footnote{The uncalculated order-$ v^2$
corrections to $e^+e^-\to J/\psi + \eta_{c2}$ generally also lead to
non-vanishing amplitudes for those helicity states other than $(\pm
1, 0)$, which are potentially as important as the ${\cal
O}(\alpha_s)$ corrections. For the leading contributions to the
corresponding polarized cross sections, one needs also include the
interference effect between the radiative and relativistic
corrections.}. We will follow this viewpoint in presenting our NLO
predictions.

In the numerical analysis, we take $\sqrt{s}=10.58$ GeV, and the
charm quark pole mass $m_c=1.4\pm 0.2$ GeV. The fine structure
constant is chosen as $\alpha(\sqrt{s}) =
1/130.9$~\cite{Bodwin:2007ga}. The running QCD strong coupling
constant is evaluated by using the two-loop formula with
$\Lambda^{(4)}_{\overline{\rm MS}}= 0.338$
GeV~\cite{Zhang:2005cha,Gong:2007db}. The nonpertubative input
parameters, {\it i.e.}, (the derivatives of ) the wave function at
the origin for $J/\psi$, $\chi^\prime_{c1}$, and $\eta_{c2}$, also
suffer from a large amount of uncertainty. Their values have been
compiled in Ref.~\cite{Eichten:1995}, which were deduced from
several different potential models. We choose to use those given by
the Buchm\"{u}ller-Tye potential model~\cite{Buchmuller:1981}:
$\vert R_{J/\psi}(0) \vert^2= 0.81 \; {\rm GeV}^3$, $\vert
R'_{\chi'_{c1}}(0) \vert^2= 0.102\;{\rm GeV}^5$, and $\vert
R''_{\eta_{c2}}(0)\vert^2 = 0.015 \;{\rm GeV}^7$.

Another important source of uncertainty for the NLO predictions
stems from the scale affiliated with the strong coupling constant.
As is well known, the scale ambiguity is a typical nuisance of NRQCD
factorization approach, reflecting the fact that two disparate hard
scales, $\sqrt{s}$ and $m_c$, are entangled together in NRQCD
short-distance coefficients. In fact, the lesson gained from the
light-cone approach strongly suggests that it is rather implausible
to set the scales entering all $\alpha_s$ in NLO short-distance
coefficient to be unanimously around $\sqrt{s}$~\cite{Jia:2010fw}.
Although we are unable to circumvent the scale ambiguity problem in
the confine of NRQCD approach, we may allow the $\mu$ to float in the range
between $2m_c$ and $\sqrt{s}$, hoping that the most trustworthy
prediction may interpolate in between.

\begin{figure}[tb]
\begin{center}
\includegraphics[scale=0.7]{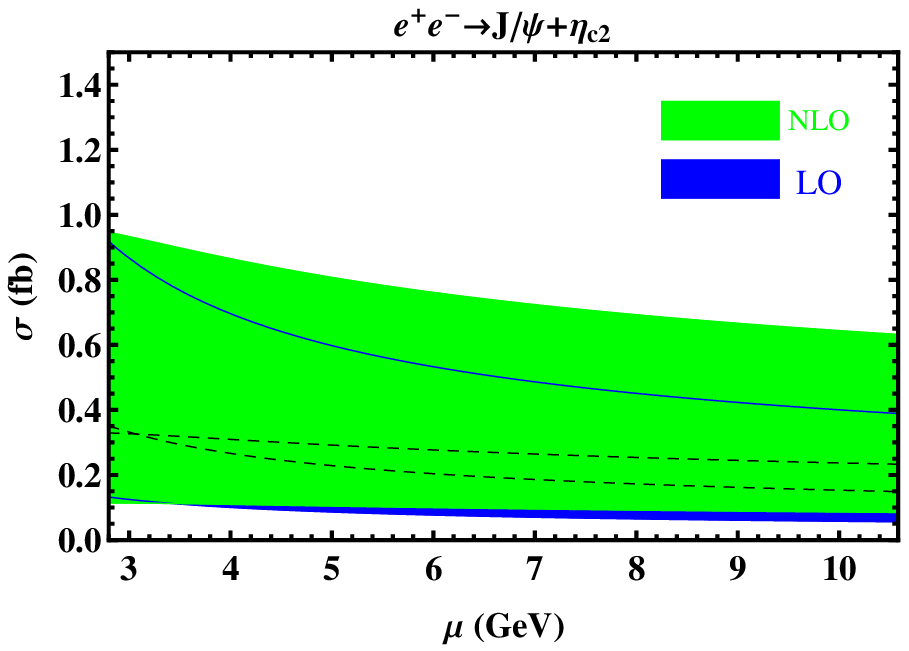}
\includegraphics[scale=0.7]{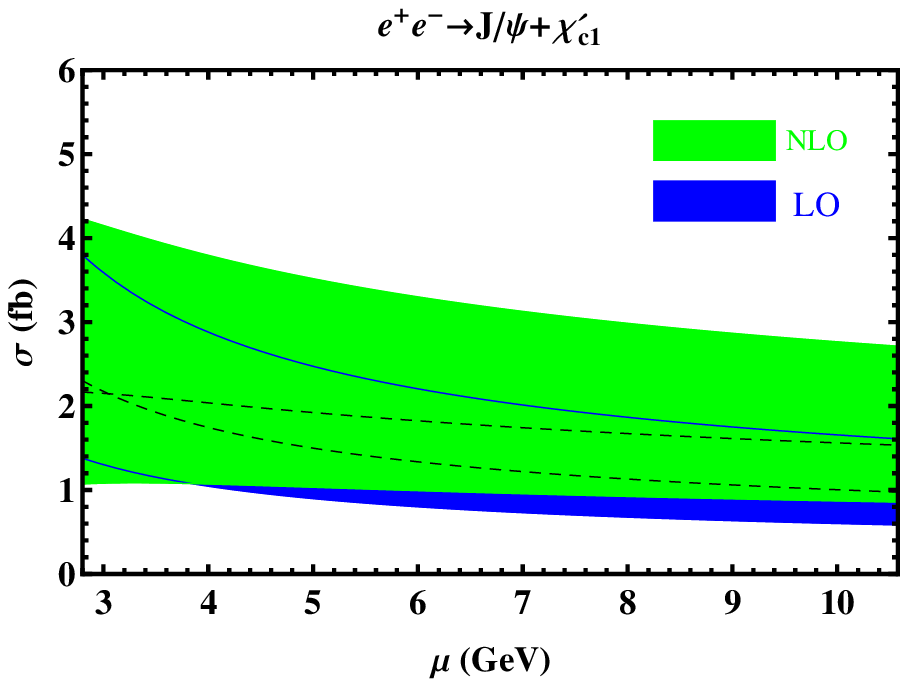}
\caption{The renormalization scale dependence of the LO and NLO
cross sections for $e^+ e^-\to J/\psi+\eta_{c2}$ (left panel) and
$e^+ e^-\to J/\psi+\chi'_{c1}$ (right panel) at $\sqrt{s} = 10.58$
GeV. The band is obtained by varying the charm quark mass in the
range $m_c=1.4 \pm 0.2$ GeV.
\label{Plot:Xection:variation:mu:error:band}}
\end{center}
\end{figure}

\begin{figure}[tb]
\begin{center}
\includegraphics[scale=0.7]{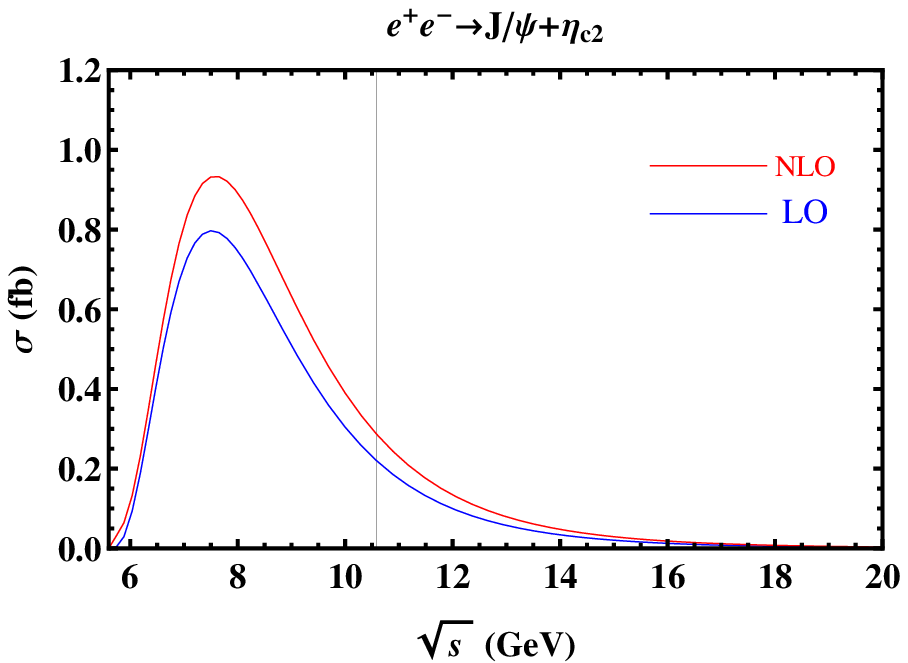}
\includegraphics[scale=0.7]{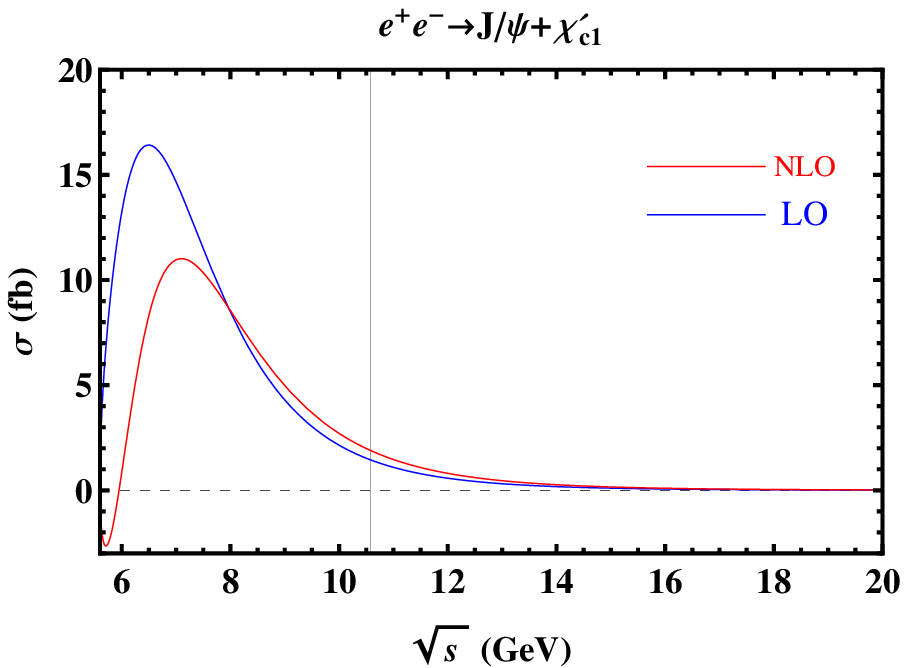}
\caption{Variation of the LO and NLO cross sections for $e^+ e^-\to
J/\psi+\eta_{c2}$ (left panel) and $e^+ e^-\to J/\psi+\chi'_{c1}$ (
right panel) with $\sqrt{s}$. $m_c$ is fixed as 1.4 GeV and the
renormalization scale is taken as $\mu={\sqrt{s}\over 2}$.
\label{Plot:Xection:variation:sqrts}}
\end{center}
\end{figure}

\begin{table*}[tb]
\caption{Polarized and total cross sections (in unit of fb) for
$e^+e^-\to J/\psi+\eta_{c2}$. We have taken $m_c=1.4$ GeV. In the
rightmost column, we also give the K factor for the unpolarized
cross sections.}
\begin{center}
\renewcommand\arraystretch{1.5}
\begin{tabular}
{>{\centering}p{0.11\textwidth}|>{\centering}p{0.11\textwidth}
>{\centering}p{0.11\textwidth}>{\centering}p{0.13\textwidth}>{\centering}p{0.13\textwidth}
>{\centering}p{0.13\textwidth}>{\centering}p{0.10\textwidth}>{\hfill}p{0.09\textwidth}<{\hfill\hfill}}
\hlinew{1pt}
   &  &
    $\sigma_{(1,0)}$   &  $\sigma_{(0,1)}$
    &  $\sigma_{(1,1)}$ & $\sigma_{(1,2)}$
     &   $\sigma_{\rm tot}$& K  \\
\hline \multirow{2}{*}{$\mu = 2 m_c$}
 & LO    & $0.18$ &    --    &    --
 &     --  & $0.35$ &
 \multirow{2}{*}{$0.94$}\\
 & NLO      & $0.17$ &    $3.4\times10^{-3}$    &    $5.5\times10^{-4}$   &
 $3.7\times10^{-5}$  &  $0.33$& \\
\hline \multirow{2}{*}{$\mu = \displaystyle\frac{\sqrt{s}}{2}$}\ \
 & LO        & $0.11$ &    --    &    --
 &     --  & $0.22$ &\multirow{2}{*}{$1.30$}\\
 & NLO     & $0.14$ &    $1.5\times10^{-3}$    &    $2.4\times10^{-4}$   &
 $1.6\times10^{-5}$  &  $0.29$& \\
\hline

\multirow{2}{*}{$\mu = \sqrt{s}$}
 & LO     & $7.4\times10^{-2}$ &    --    &    --
 &     --  & $0.15$ &\multirow{2}{*}{$1.57$}\\
 & NLO   & $0.12$ &    $6.8\times10^{-4}$    &    $1.1\times10^{-4}$
 &   $7.5\times10^{-6}$  &  $0.23$& \\
\hline
\hlinew{1pt}
\end{tabular}
\end{center}
\label{Table-Xsection:jpsi:etac2}
\end{table*}


\begin{table*}[htb]
\caption{Polarized and total cross sections (in unit of fb) for
$e^+e^-\to J/\psi+\chi'_{c1}$. We have taken $m_c=1.4$ GeV. In the
rightmost column, we also list the K factor for the unpolarized
cross sections.}
\begin{center}
\renewcommand\arraystretch{1.5}
\begin{tabular}
{>{\centering}p{0.13\textwidth}|>{\centering}p{0.13\textwidth}
>{\centering}p{0.13\textwidth}>{\centering}p{0.13\textwidth}>{\centering}p{0.13\textwidth}
>{\centering}p{0.13\textwidth}>{\hfill}p{0.13\textwidth}<{\hfill\hfill}}
\hlinew{1pt}
 &  &
    $\sigma_{(1,0)}$   &  $\sigma_{(0,1)}$
    &  $\sigma_{(1,1)}$ &   $\sigma_{\rm tot}$& K  \\
\hline
\multirow{2}{*}{$\mu = 2 m_c$}
 & LO & $2.3\times10^{-3}$ &    $1.07$    &    $0.082$  & $2.30$ &
 \multirow{2}{*}{$0.94$}\\
 & NLO  & $-0.026$ &    $1.08$    &    $0.033$   &  $2.16$& \\
\hline
\multirow{2}{*}{$\mu = \displaystyle\frac{\sqrt{s}}{2}$}\ \
 & LO  & $1.4\times10^{-3}$ &    $0.67$    &    $0.051$ & $1.44$ &\multirow{2}{*}{$1.31$} \\
 & NLO & $-0.012$ &    $0.91$    &    $0.045$  &  $1.89$& \\
\hline

\multirow{2}{*}{$\mu = \sqrt{s}$}
 & LO   & $9.7\times10^{-4}$ &   $0.45$    &   0.035  & $0.98$ &\multirow{2}{*}{$1.57$}\\
 & NLO   & $-0.0063$ &    $0.73$    &    $0.042$  &  $1.53$& \\
\hline
\hlinew{1pt}
\end{tabular}
\end{center}
\label{Table-Xsection:jpsi:chic1prime}
\end{table*}

In Fig.~\ref{Plot:Xection:variation:mu:error:band}, we show the
total cross sections both for $e^+ e^-\to J/\psi+\eta_{c2}$  and
$e^+ e^-\to J/\psi+\chi'_{c1}$ as a function of $\mu$, which also
take into account the error due to the uncertainty of $m_c$. One
sees clearly that, after incorporating the NLO perturbative
correction, the scale dependence of the cross section has been
reduced. In Fig.~\ref{Plot:Xection:variation:sqrts}, we also plot
the LO and NLO total cross sections for $e^+ e^-\to
J/\psi+\eta_{c2}$ as a function of $\sqrt{s}$. Since the
leading-twist $(0,0)$ channel is absent in both of the processes,
the cross sections drop very rapidly ($\propto 1/s^4$) as $\sqrt{s}$
increases.

In Tables~\ref{Table-Xsection:jpsi:etac2} and
\ref{Table-Xsection:jpsi:chic1prime}, we also list the predictions
for the polarized cross sections from each individual helicity
channel as well as the total cross sections for $e^+e^-\to
J/\psi+\eta_{c2}$ and $e^+e^-\to J/\psi+\chi^\prime_{c1}$, with
$m_c$ fixed at 1.4 GeV but with the renormalization scale chosen at
three different points. The impact of the ${\cal O}(\alpha_s)$
corrections to the total cross sections is surprisingly alike for
both double-charmonium production processes: incorporating the NLO
pertubative correction decreases the LO cross sections by about 6\%
for $\mu=2 m_c$, or increases the LO result by roughly 30\% for
$\mu=\sqrt{s}/2$, or enhances the LO result by about 60\% for
$\mu=\sqrt{s}$. Hence, the relative importance of the ${\cal
O}(\alpha_s)$ correction increases as $\mu$ increases. Taking the
medium value $\mu=\sqrt{s}/2$, our NLO predictions for the cross
sections of $e^+e^-\to J/\psi+\eta_{c2}$ and $e^+e^-\to
J/\psi+\chi^\prime_{c1}$ at $\sqrt{s}=10.58$ GeV reach about 0.3 fb
and 1.9 fb, respectively. The production rate of the latter process
is about six times more copious than that of the former.

Inspecting Tables~\ref{Table-Xsection:jpsi:etac2} and
\ref{Table-Xsection:jpsi:chic1prime}, one observes an interesting
hierarchy of the polarized cross sections for the above processes.
The overwhelming contribution for the $J/\psi+\eta_{c2}$ production
comes solely from the $(\pm 1,0)$ state, with $J/\psi$ transversely
polarized, while the dominant helicity channel for producing
$J/\psi+\chi'_{c1}$ is instead the $(0,\pm 1)$ state, with $J/\psi$
longitudinally polarized. This characteristic of $J/\psi$
polarization may serve as a benchmark for the future
double-charmonium production experiment to unveil the nature of the
$X(3872)$, provided that the prospective Super $B$ factory can
observe sufficient number of $X(3872)$ events recoiling against
$J/\psi$.

We can ascertain the observation potential of these two exclusive
double charmonium production processes. Thus far, \textsc{Belle}
experiment has accumulated about $1000\;{\rm fb}^{-1}$ data. Taking
$\sigma[e^+e^-\to J/\psi+\eta_{c2}]\approx 0.23-0.33$ and
$\sigma[e^+e^-\to J/\psi+\chi^\prime_{c1}]\approx 1.53-2.16$ fb from
Tables~\ref{Table-Xsection:jpsi:etac2} and
\ref{Table-Xsection:jpsi:chic1prime}, we estimate that roughly
$230-330$ $J/\psi+\eta_{c2}$ events, and  $1500-2200$
$J/\psi+\chi'_{c1}$ events have been produced at \textsc{Belle} at
around $\sqrt{s}=10.58$ GeV.

Since neither of the masses and decay patterns of $\eta_{c2}$ and
$\chi^\prime_{c1}$ is known, it seems experimentally impossible to
simultaneously reconstruct the $J/\psi$ and the
$\eta_{c2}(\chi^\prime_{c1})$ signals. The viable experimental way
is to only reconstruct the $J/\psi\to l^+l^-$ event, and fit the
recoil mass spectrum against the $J/\psi$ to estimate the number of
$\eta_{c2}(\chi'_{c1})$ peak events. This method does not depend on
the concrete decay modes of $\eta_{c2}(\chi'_{c1})$, and is
particularly suitable for the limited statistics of signal events
like in our case. In fact, this method has already been used
routinely by the \textsc{Belle} collaboration to search for the
double charmonium production processes $e^+e^- \to J/\psi+C$-even
charmonium.

In fitting the recoil mass spectrum, the net detection efficiency
for $e^+e^-\to J/\psi+ \eta_{c2}(\chi'_{c1})$ may reach around 4\%
(Similar for $\eta_{c2}$ and $\chi'_{c1}$, with the reconstruction
efficiency for $J/\psi\to l^+l^-$
included~\cite{Shen:2012:private}). As a very crude estimate, the
number of observed $e^+e^-\to J/\psi+ \eta_{c2}(\chi'_{c1})$ events
are expected to reach $(230-330)\times 4\%=9-13$, and
$(1500-2200)\times 4\%=60-90$, respectively. Since only the $J/\psi$
is reconstructed, the background level can be a little higher. With
only around $9-13$ observed $J/\psi+\eta_{c2}$ events, it appears
difficult to observe a significant signal with current  1 ${\rm
ab}^{-1}$ \textsc{Belle} data. If $\chi^\prime_{c1}$ is indeed the
$X(3872)$ meson, thanks to the very narrow width of the $X(3872)$,
it seems possible to observe the 60-90 $J/\psi+\chi'_{c1}$ signal
events with the current \textsc{Belle} full data sample, which may
provide a strong incentive for updating their earlier $e^+ e^-\to
J/\psi +$ charmonium analysis with only $673$ ${\rm fb}^{-1}$
data~\cite{Abe:2007jn,Pakhlov:2009nj}. This study may hopefully help
one to better understand the mechanism about the $X(3872)$
production. On the other hand, in case that $\chi^\prime_{c1}$ is
not $X(3872)$ and its width is no longer narrow, it may be difficult
to observe a significant signal at the moment.

In any event, a much larger data samples seem to be called for to
arrive at some definite conclusion. In the prospective Super $B$
factory, which may reach an integrated luminosity of $50\;{\rm
ab}^{-1}$ by 2022, it seems feasible that the processes $e^+e^-\to
J/\psi+\eta_{c2}(\chi'_{c1})$ will eventually be observed, and the
quantum number of $X(3872)$ also hopefully be nailed down.

\section{Summary}
\label{summary}

In this work we have calculated the complete NLO perturbative
corrections to $e^+e^-\to J/\psi+\eta_{c2}$ within the NRQCD
factorization framework, and found that the ${\cal O}(\alpha_s)$
correction can be sizable for a larger value of the renormalization
scale. Our calculation indicates that the dominate contribution to
the total cross section comes from the helicity states $(\pm 1,0)$.

We have also carried out a comparative study for $e^+e^-\to
J/\psi+\eta_{c2}$ and $e^+e^-\to J/\psi+\chi^\prime_{c1}$ at $B$
factory energy, with the specific motivation that these double
charmonium production processes may provide useful means to unveil
the quantum number of the $X(3872)$ meson in the future $B$
factories. It turns out that the production rate of the latter
process is about 6-7 times greater than the former, therefore it is
possible to observe the $J/\psi+\chi^\prime_{c1}$ signals based on
the current 1 ${\rm ab}^{-1}$ \textsc{Belle} data sample, if the
$\chi^\prime_{c1}$ is indeed the very narrow $X(3872)$ particle. The
dominantly transverse polarization of the $J/\psi$ is also a useful
indicator for identifying the $X(3872)$ with $\chi^\prime_{c1}$.

A necessary extension of our current work is to further incorporate
the leading relativistic correction to the above double-charmonium
production processes, whose effects might be as important as the
radiative corrections. The study along this direction is of some
theoretical interest, especially regarding that, until today the
${\cal O}(v^2)$ calculation hardly exists for the production
processes involving the $P$, $D$-wave quarkonium.

It has been recently realized that the ${\cal O}(\alpha_s)$ NRQCD
short-distance coefficients for the helicity-suppressed double
charmonium-production processes are often plagued with the double
logarithm of form $\ln^2(s/m_c^2)$. This symptom is likely
intertwined with some long-standing failure of applying the
light-cone approach to the ``higher-twist" hard exclusive reactions
beyond tree level. To expedite a better understanding and
controlling of these double logarithms, we have scrutinized all the
NLO diagrams for the various exclusive double-charmonium processes
that have been studied so far, {\it e.g.}, $e^+ e^- \to J/\psi+
\eta_{c2}(\eta_c,\chi_{c0,1,2})$, and singled out those that contain
the double logarithm, and enumerate their coefficients for each
individual helicity states. No simple and general pattern has been
recognized yet.

\begin{acknowledgments}
We thank Wen-Long Sang for checking some part of our results.
We also thank Cheng-Ping Shen for helpful discussions on the
experimental issues.
This research was supported in part by the National Natural Science
Foundation of China under Grant Nos.~10935012, 11125525, DFG and
NSFC (CRC 110), and by the Ministry of Science and Technology of
China under Contract No. 2009CB825200.
\end{acknowledgments}

\appendix

\section{Anatomy of the NLO diagrams containing double
logarithm in various double-charmonium production processes}%
\label{double:log:anatomy}

\begin{figure}
\begin{center}
\includegraphics[height=9 cm]{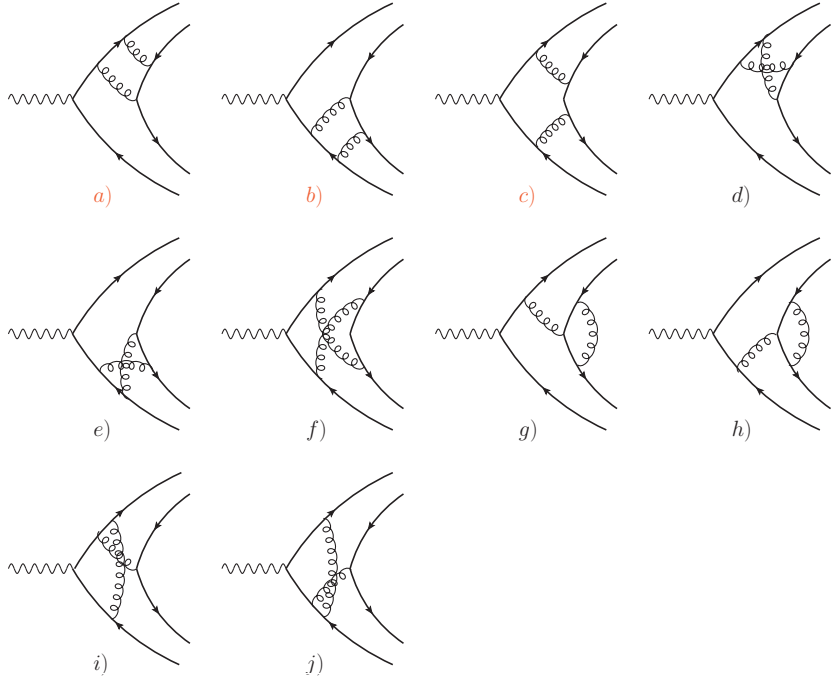}
\caption{The NLO diagrams that contain the double logarithm
$\ln^2(s/m_c^2)$ (in Feynman gauge) for the process $\gamma^*\to
J/\psi+\eta_{c2}\,(\eta_c,\chi_{c0,1,2})$. The respective
charge-conjugated diagrams have been suppressed.
\label{Fig:diagram:jpsi:etac:double:log}}
\end{center}
\end{figure}

Although the NRQCD factorization approach and the light-cone
approach are based on two completely different expansion strategies,
they can be intimately linked for the exclusive double charmonium
production processes in the limit $s \gg  m_c^2$. As exemplified by
an anatomy of the ${\cal O}(\alpha_s)$ correction to the $B_c$
electromagnetic form factor~\cite{Jia:2010fw}, the light-cone
approach can be efficiently utilized to reproduce the asymptotic
behavior of the ${\cal O}(\alpha_s)$ NRQCD short-distance
coefficient through the idea of
refactorization~\cite{Ma:2006hc,Bell:2008er,Jia:2008ep}.

It is worth emphasizing one important limitation of the
refactorization procedure. So far it can only be successfully
applied to the ``leading-twist" reactions involving quarkonium, that
is, with the HSR-favored helicity configurations. A general
characteristic about this class of of processes is that, at the NLO
in $\alpha_s$, only the single collinear logarithm of $\ln{s\over
m_c^2}$ arises in the NRQCD short-distance coefficient, which can be
resummed to all orders in $\alpha_s$ with the aid of the
Brodsky-Lepage evolution equation. This feature has been explicitly
verified in several examples, {\it e.g.}, $\gamma^*\to
\eta_c+\gamma$~\cite{Shifman:1980dk,Jia:2008ep,Sang:2009jc},
$\gamma^*+B_c\to B_c$~\cite{Jia:2008ep}, $\gamma^*\rightarrow
J/\psi(0)+\chi_{c0,2}(0)$~\cite{Dong:2011fb}.

However, the majority of phenomenologically relevant
double-charmonia production processes are of ``higher-twist" nature,
exemplified by $\gamma^*\to J/\psi+\eta_{c2}(\eta_c,\chi_{c0,1,2})$,
for which some of or all of possible helicity channels are of
HSR-suppressed type. Until today, it is still not clear how to
consistently compute the NLO radiative correction to this type of
processes in the light-cone framework, due to some long-standing
problem such as the inevitable emergence of the endpoint
singularity. By contrast, NRQCD factorization approach, which is
based on the velocity expansion rather than the twist expansion,
serves as the {\it only} viable tool to investigate the ${\cal
O}(\alpha_s)$ corrections to this kind of helicity-suppressed hard
exclusive reactions.

An abnormal feature about the ${\cal O}(\alpha_s)$ NRQCD
short-distance coefficients for the helicity-suppressed double
charmonium production processes seems to be the frequent emergence
of the double logarithm $\ln^2(s/m_c^2)$, which seems to seriously
deteriorate the convergence of the perturbative expansion in NRQCD
factorization~\footnote{In fact, one will encounter the double
logarithm even in the single-quarkonium exclusive production process
such as $\gamma^*\to \eta_c+\gamma$, once the corresponding ${\cal
O}(\alpha_s)$ NRQCD short-distance coefficient is expanded to NLO in
$m_c^2/s$. This can be readily checked by using the analytic ${\cal
O}(\alpha_s)$ short-distance coefficient given in
\cite{Shifman:1980dk,Sang:2009jc}.}. This double logarithm appears
to result from the overlap between the ``collinear" and ``soft"
regions in the NRQCD short-distance coefficient and seems deeply
related to the afore-mentioned endpoint singularity problem.
Although it is feasible to reproduce the closed form of the double
logarithms for each concrete process, the lack of a thorough
understanding of their behavior in higher order in $\alpha_s$
prevents one from systematically controlling them, {\it i.e.}, to
resum them to all orders in $\alpha_s$.

A useful starting point is to anatomize all the double logarithms of
form $\ln^2(s/m_c^2)$ from the existing NLO calculations for the
various exclusive double charmonium processes. For this purpose, we
examine all the NLO Feynman diagrams for the processes $e^+ e^- \to
J/\psi+ \eta_{c2}(\eta_c,\chi_{c0,1,2})$, sort out those that
contain the double logarithm, and enumerate their coefficients.

A careful examination reveals that (almost) all the relevant NLO
diagrams that contribute the double logarithm $\ln^2(s/m_c^2)$ (in
Feynman gauge) for the above processes are included in
Fig.~\ref{Fig:diagram:jpsi:etac:double:log}, which was first
identified in \cite{Jia:2010fw}.

\begin{table*}[htb]
\caption{The coefficients of the double logarithm associated with
each diagram in Fig.~\ref{Fig:diagram:jpsi:etac:double:log} from the
various helicity states $(\lambda_1, \lambda_2)$ in the processes
$\gamma^* \to J/\psi + \eta_{c2}(\eta_c)$. The two types of color
factors are denoted by $C_1 = C_F^2$ and $C_2 = C_F(C_F- {1\over
2}C_A)$, where $C_F={4 \over 3}$ and $C_A=3$.}
\begin{center}
\renewcommand\arraystretch{2.0}
\begin{tabular}{|>{\centering}p{0.07\textwidth}|>{\centering}p{0.07\textwidth}|>{\centering}p{0.07\textwidth}|>{\centering}p{0.07\textwidth}|>{\centering}p{0.07\textwidth}
|>{\centering}p{0.07\textwidth}|>{\centering}p{0.07\textwidth}|>{\centering}p{0.07\textwidth}|>{\centering}p{0.07\textwidth}|>{\centering}p{0.07\textwidth}
|>{\centering}p{0.07\textwidth}|>{\hfill}p{0.07\textwidth}<{\hfill\hfill}|}
\hline
  \multicolumn{2}{|c|}{\rm Diagrams} & $a)$ & $b)$ & $c)$ & $d)$ & $e)$ & $f)$ & $g)$ & $h)$ & $i)$ &
  $j)$
\\\hline
 \multicolumn{2}{|c|} {\rm Color Factor} & $C_1$ & $C_1$ & $C_1$ & $C_2$ & $C_2$ & $C_2$ & $C_2$ & $C_2$
 & $C_2$ & $C_2$
\\\hline
  \multirow{4}{*}{$\eta_{c2}$}
  & (1,0) & $\frac{3}{64}$ & $\frac{9}{128}$ & $\frac{3}{256}$ & $\frac{9}{64}$ & $\frac{3}{128}$ & $\frac{21}{128}$ & $-\frac{9}{128}$ & $-\frac{3}{128}$ & $-\frac{9}{128}$ & $-\frac{3}{128}$
\\
  & (0,1) & -- & $-\frac{3\sqrt{3}}{128}$ & -- & -- & -- & $-\frac{3\sqrt{3}}{128}$ & -- & -- & -- & --
\\
  & (1,1) & -- & $-\frac{3\sqrt{3}}{256}$ & $\frac{3\sqrt{3}}{256}$ & -- & -- & -- & -- & -- & -- & --
\\
  & (1,2) & -- & $-\frac{3\sqrt{6}}{256}$ & -- & -- & -- & -- & -- & -- & -- & --
\\ \hline
$\eta_c$ & (1,0) & $\frac{3}{64}$ & $\frac{3}{32}$ & $\frac{3}{128}$ & $\frac{3}{32}$ & $\frac{3}{64}$ & $\frac{9}{64}$ & $-\frac{3}{64}$ & $-\frac{3}{64}$ & $-\frac{3}{64}$ & $-\frac{3}{64}$
\\ \hline
\end{tabular}
\end{center}
\label{Table:double:logarithm:etac2:etac}
\end{table*}

In Table~\ref{Table:double:logarithm:etac2:etac}, we tabulate the
double logarithms that come from each individual diagram in
Fig.~\ref{Fig:diagram:jpsi:etac:double:log}, with the
color-structure specified, for the processes $e^+ e^- \to J/\psi+
\eta_{c2}(\eta_c)$.

The reaction $e^+ e^- \to J/\psi(\lambda_1)+ \eta_c(\lambda_2)$ only
possesses one independent helicity configuration $(\pm 1,0)$. Each
diagram contains a non-vanishing double logarithm. Summing up all
their contributions, and multiply by 2 to account for the
contribution from the charge-conjugated diagrams, we recover the net
double logarithm in the NLO short-distance coefficient:
${C^{(1)}\over C^{(0)}}\big|_{\rm asym}={13\over 24}\ln^2{s\over
m_c^2}$, as given in equation (37) of Ref.~\cite{Jia:2010fw}.

All the four polarized production channels $e^+e^- \to
J/\psi(\lambda_1)+\eta_{c2}(\lambda_2)$ are helicity-suppressed.
Adding up the contributions from all the diagrams in
Fig.~\ref{Fig:diagram:jpsi:etac:double:log}, and multiply by 2, we
recover the net double logarithm in the NLO reduced helicity
amplitudes $a^{(1)}_{\lambda_1,\lambda_2}\big|_{\rm asym}$, as given
in equations (\ref{red:ampl:asymp:NLO:Jpsi:etac2}). Note that the
double logarithms from the $(\pm 1,0)$ channel has the similar
diagram-by-diagram structure as its counterpart for $e^+e^- \to
J/\psi+\eta_{c}$. Nevertheless, for the remaining three helicity
channels, a great simplification arises that many of the diagrams in
Fig.~\ref{Fig:diagram:jpsi:etac:double:log} do not contain double
logarithm.

\begin{figure}
\begin{center}
\includegraphics[height=4 cm]{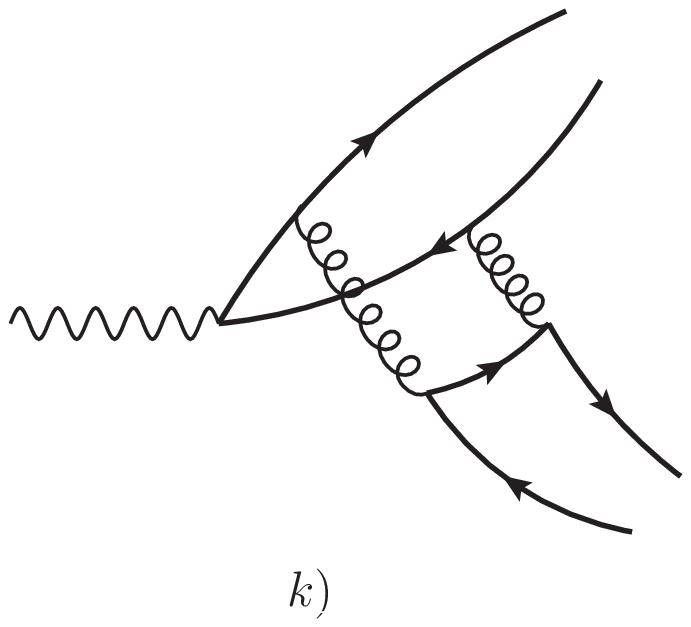}
\caption{The extra NLO diagram relative to
Fig.~\ref{Fig:diagram:jpsi:etac:double:log} that also contributes
the double logarithm $\ln^2(s/m_c^2)$ (in Feynman gauge) to the
helicity channel $(\pm 1,\pm2)$ for $\gamma^*\to J/\psi+\chi_{c2}$.
The respective charge-conjugated diagram is not shown.
\label{Fig:diagram:jpsi:chic2:12:extra:double:log}}
\end{center}
\end{figure}

\begin{table*}[htb]
\caption{The coefficients of the double logarithm associated with
each diagram in Fig.~\ref{Fig:diagram:jpsi:etac:double:log} from the
various helicity states $(\lambda_1, \lambda_2)$ in the process
$\gamma^* \to J/\psi + \chi_{c0,1,2}$. The two types of color
factors are represented by $C_1 = C_F^2$ and $C_2 = C_F(C_F- {1\over
2}C_A)$. }
\begin{center}
\renewcommand\arraystretch{2.0}
\begin{tabular}{|>{\centering}p{0.07\textwidth}|>{\centering}p{0.07\textwidth}|
>{\centering}p{0.07\textwidth}|>{\centering}p{0.07\textwidth}|>{\centering}p{0.07\textwidth}
|>{\centering}p{0.07\textwidth}|>{\centering}p{0.07\textwidth}|>{\centering}p{0.07\textwidth}
|>{\centering}p{0.07\textwidth}|>{\centering}p{0.07\textwidth}
|>{\centering}p{0.07\textwidth}|>{\hfill}p{0.07\textwidth}<{\hfill\hfill}|}
\hline
  \multicolumn{2}{|c|}{\rm Diagrams} & $a)$ & $b)$ & $c)$ & $d)$ & $e)$ & $f)$ & $g)$ & $h)$ & $i)$ &
  $j)$
\\\hline
 \multicolumn{2}{|c|}{\rm Color Factor} & $C_1$ & $C_1$ & $C_1$ & $C_2$
 & $C_2$ & $C_2$ & $C_2$ & $C_2$ & $C_2$ & $C_2$
\\\hline
 $\chi_{c0}$ & (1,0) & $\frac{7}{384}$ & $\frac{1}{192}$ & $\frac{1}{768}$ & $\frac{1}{96}$ & $\frac{7}{384}$ & $\frac{11}{384}$ & $-\frac{1}{192}$ & $-\frac{7}{384}$ & $-\frac{1}{192}$ & $-\frac{7}{384}$
\\\hline
 \multirow{3}{*}{$\chi_{c1}$} & (1,0) & -- & $-\frac{3}{128}$ & $-\frac{3}{256}$ & $-\frac{3}{64}$ & $\frac{3}{128}$ & $-\frac{3}{128}$ & $\frac{3}{128}$ & $-\frac{3}{128}$ & $\frac{3}{128}$ & $-\frac{3}{128}$
\\
   & (0,1) & $\frac{9}{512}$ & $\frac{3}{256}$ & $\frac{9}{1024}$ & $\frac{3}{512}$ & $\frac{9}{512}$ & $\frac{9}{256}$ & $-\frac{3}{512}$ & $-\frac{9}{512}$ & $-\frac{3}{512}$ & $-\frac{9}{512}$
\\
  & (1,1) & $\frac{27}{1024}$ & $\frac{3}{512}$ & $-\frac{3}{1024}$ & $-\frac{3}{512}$ & $\frac{15}{512}$ & $\frac{3}{128}$ & $\frac{3}{512}$ & $-\frac{15}{512}$ & $\frac{3}{512}$ & $-\frac{15}{512}$
\\ \hline
  \multirow{4}{*}{$\chi_{c2}$} & (0,1) & $-\frac{3}{256}$ & $-\frac{3}{128}$ & $-\frac{3}{512}$ & $-\frac{3}{256}$ & $-\frac{3}{256}$ & $-\frac{3}{64}$ & $\frac{3}{256}$ & $\frac{3}{256}$ & $\frac{3}{256}$ & $\frac{3}{256}$
\\
  & (1,0) & $-\frac{3}{1408}$ & $-\frac{3}{704}$ & $-\frac{3}{2816}$ & $-\frac{3}{352}$ & $-\frac{3}{1408}$ & $-\frac{15}{1408}$ & $\frac{3}{704}$ & $\frac{3}{1408}$ & $\frac{3}{704}$ & $\frac{3}{1408}$
\\
  & (1,1) & $\frac{9}{1024}$ & $\frac{9}{512}$ & $-\frac{9}{1024}$ & $\frac{15}{512}$ & $\frac{3}{512}$ & $\frac{15}{256}$ & $-\frac{9}{512}$ & $-\frac{3}{512}$ & $-\frac{9}{512}$ & $-\frac{3}{512}$
\\
  & (1,2) & -- & $\frac{3}{256}$ & $\frac{3}{256}$ & $\frac{3}{128}$ & -- & $\frac{9}{128}$ & $-\frac{3}{128}$ & -- & $-\frac{3}{128}$ & --
\\ \hline
\end{tabular}
\end{center}
\label{Table:double:logarithm:chic0:1:2}
\end{table*}

Next we examine the double logarithms that are affiliated with the
polarized production processes $e^+e^- \to
J/\psi(\lambda_1)+\chi_{cJ}(\lambda_2)$ ($J=0,1,2$). Excluding two
helicity-conserved channels $\gamma^* \to J/\psi(0)+\chi_{c0,2}(0)$,
Table~\ref{Table:double:logarithm:chic0:1:2} lists the structure of
double logarithms in the remaining eight helicity-suppressed
channels. A noteworthy thing is that, for the helicity channel $(\pm
1,\pm 2)$ of $\gamma^*\to J/\psi+\chi_{c2}$, besides the diagrams
shown in Fig.~\ref{Fig:diagram:jpsi:etac:double:log}, one extra NLO
diagram as shown in
Fig.~\ref{Fig:diagram:jpsi:chic2:12:extra:double:log} also need be
included. The color factor associated with the diagram in
Fig.~\ref{Fig:diagram:jpsi:chic2:12:extra:double:log} is ${1 \over
2} C_F$, and the corresponding coefficient is $-{9 \over 64}$.

Adding up the contributions from all the diagrams in
Fig.~\ref{Fig:diagram:jpsi:etac:double:log} for the reactions
$e^+e^- \to J/\psi(\lambda_1)+\chi_{c0,1,2}(\lambda_2)$, together
with that in Fig.~\ref{Fig:diagram:jpsi:chic2:12:extra:double:log}
for the $(\pm 1,\pm 2)$ state, and multiply the results by 2, we
then reproduce the net double logarithms in the NLO reduced helicity
amplitudes, $K^J_{\lambda_1,\lambda_2}(r,{\mu^2\over s})_{\rm asym}$
given in equations (14) through (16) in Ref.~\cite{Dong:2011fb}.

An obvious observation from
Tables~\ref{Table:double:logarithm:etac2:etac} and
\ref{Table:double:logarithm:chic0:1:2} is that the specific
structures of the double logarithms are process-dependent. It seems
that no simple, unified pattern can be readily recognized.

\end{document}